# The excitation of a quantum gas of independent particles under periodic perturbation in integrable or non-integrable potential [*]

P. Magierski[1,2], J. Skalski[3], J. Blocki[3]

[1] *The Royal Institute of Technology, Physics Department Frescati,
Frescativägen 24, S-10405 Stockholm, Sweden*

[2] *Institute of Physics, Warsaw University of Technology,
ul. Koszykowa 75, PL-00662 Warsaw, Poland*

[3] *Soltan Institute for Nuclear Studies,
Hoza 69, 00 681 Warsaw, Poland*

(March 25, 1997)

## Abstract

The excitation of a quantum gas of 112 independent fermions in the time-dependent potential well, periodically oscillating around the spherical shape, was followed over 10 oscillation cycles. Five different oscillation frequencies are considered for each of the five types of deformations: spheroidal and Legendre polynomial ripples: $P_3$, $P_4$, $P_5$ and $P_6$. The excitation rate of the gas in the deforming hard-walled cavities substantially decreases after initial one or two cycles and the final excitation energy is a few times smaller than the wall formula predictions. Qualitatively similar results are obtained for the diffused Woods-Saxon well. The details and possible origins of this behaviour are discussed as well as the consequences for the one-body dissipation model of nuclear dynamics.

Typeset using REVTeX

# I. INTRODUCTION

In this paper we study the excitation of a quantum system of independent fermions in a container periodically changing its shape. This work may be treated as a continuation of studies reported in [1,2] and aims at better understanding of the one-body dissipation mechanism, pertinent to the nuclear dynamics.

The important role of dissipative effects in nuclear dynamics, i.e. of energy transfer from a few collective to many non-collective (chaotic) degrees of freedom, has been confirmed by numerous experiments on heavy ion collisions in which deep inelastic scattering, fission and fusion processes were studied (see e.g. [3–5] and references quoted therein). Two mechanisms of nuclear friction are usually suggested: two-body collisions and the collisions of (quasi-)nucleons with the nuclear mean field. The former resembles the dissipative effects in fluid dynamics, arising from viscous shearing stress between adjacent layers of fluid in nonuniform motion, and is associated with a relatively short mean free path of nucleons. The latter mechanism, called one-body dissipation, operates due to the Pauli principle which inhibits two-nucleon scattering, and is suggested by the experimental evidence for a long mean free path of nucleons in nuclei close to their ground states. Which of these two mechanisms prevails in a specific energy range is still an unsolved problem but it is believed that, for sufficiently low energies, the one-body dissipation should dominate.

The question we address here is the dissipative response of the studied system during 10 oscillation periods, i.e. the oscillation time ten times longer than the one studied in Refs. [1,2]. While these previous results were in fact related more to non-cyclic processes, like fission, the periodicity of driving becomes vital in the present case. Obtained results shed some light on a number of questions.

First, we compare the behaviour of the quantal gas in the vibrating cavity with the classical predictions based on the 'wall formula' [6]. The latter gives the rate of energy absorption by a gas of noninteracting particles moving chaotically inside a slowly time-dependent container. Its derivation is based on the chaotic adiabatic dynamics [7–10] in which the distinction is being made between fast (single-particle) and slow (collective) degrees of freedom and the damping of collective motion results from the coupling to the fast modes. In the classical domain, the range of validity of the wall fomula for sufficiently chaotic systems seems to extend beyond the adiabaticity constraint, as shown by numerical studies [11]. Similar conclusion follows from studies of quantum systems [1,2], in which, however, only one oscillation period was investigated. The present results enable a more thorough comparison between quantum and classical dynamics.

Secondly, under a number of assumptions validating the stochastic approach, the excitation of a quantum system of independent particles may be viewed as the diffusion process, with the independent and dependent variables being the state number and the transition probability between single-particle levels around the Fermi surface, respectively [12]. This way of description implies the irreversibility of the nuclear motion (i.e. reversing the time arrow one will not recover the initial state). In this picture, the transition probability is a gaussian, with the variance proportional to the excitation energy. The dissipation rate comes out proportional to a diffusion constant which is directly related to the time-dependent Hamiltonian. Studying the shape of the occupation probabilities we directly check the validity of this picture of excitation in the quantum system studied.

Thirdly, there is a phenomenon of saturation of excitation in periodically (externally) driven quantum systems which show unlimited energy growth in their classical version (see [13] and references quoted therein). The reason of saturation is that, loosely speaking, a general initial state usually has substantial overlaps only with a finite number of the time-recurring Floquet eigenstates (localization). It follows that the time-evolution of such an initial state is quasiperiodic an its excitation energy limited. This property seems to be quite general and *a priori* expected in a generic system. Our study allows to address the non-trivial question: What is the time scale of this saturation and how does it depend on the driving frequency in the particular system of interest for nuclear physics?

Although this work is motivated by nuclear physics phenomena it is related also to the general question of the excitation mechanism in periodically driven mesoscopic systems of non-interacting (quasi-)particles.

## II. THE MODEL

In this section we describe briefly our model. More detailed description may be found in Ref. [1]. The shape of the cavity is defined by:

$$R(\theta, t) = \frac{R_0}{\lambda(t)}[1 + \alpha_1(t)P_1(\cos\theta) + \alpha_n(t)P_n(\cos\theta)], \quad (1)$$

for Legendre polynomial distortions, and for spheroidal distortions

$$\frac{R^2 \sin^2\theta}{a^2} + \frac{R^2 \cos^2\theta}{b^2} = 1, \quad (2)$$

where

$$b = R_0[1 + \alpha(t)], \qquad a = \frac{R_0}{\sqrt{1+\alpha(t)}}. \quad (3)$$

The coefficients $\lambda(t)$ and $\alpha_1(t)$ fix the volume of the cavity and its center of mass, respectively. The time-dependent amplitudes $\alpha_n(t)$ and $\alpha(t)$ are

$$\alpha_n(t) = c\sqrt{\frac{2n+1}{5}}\cos\omega t, \qquad n = 3, 4, 5, 6, \quad (4)$$

$$\alpha(t) = c\cos\omega t \quad . \quad (5)$$

The value of $c$ was taken equal to 0.2, which in real nuclei would mean a quite sizable amplitude of vibrations, $\beta_n \approx 0.32$. The factor $\sqrt{(2n+1)/5}$ ensures approximately the same rms deviation from the spherical shape for all deformations considered.

The radius of the cavity is fixed at the value $R_0 = 1.16 \times (184)^{1/3}$fm $= 6.5978$ fm, corresponding to the nucleus with $A = 184$ nucleons. The cavity has been filled with 112 nucleons of the same kind, which can be thought of as neutrons.

The adiabaticity parameter, $\eta$, is the ratio of the highest speed attained by a surface element for $n = 2$ to a nominal Fermi velocity of an $N$-fermion gas:

$$\eta = \frac{c\omega R_0}{v_F}. \tag{6}$$

For fixed $v_F$, the adiabatic limit corresponds to $\eta \to 0$. In our model, $\omega$ and $\eta$ are related by [1]

$$\hbar\omega = 44.07\eta \quad \text{MeV}. \tag{7}$$

The time evolution of the system is followed by solving the time-dependent Schroedinger equation in hard-walled cavities, eqs. (1,2), for a number of wave functions $|\psi_n(t)\rangle$. At time $t = 0$ these wave functions are equal to the lowest eigenstates of the instantaneous Hamiltonian, $\hat{H}_0 = \hat{H}(t=0)$,

$$\hat{H}_0 \mid \phi_j \rangle = \epsilon_j \mid \phi_j \rangle. \tag{8}$$

After the integer number of periods, $t = kT$ ($k$ =1,2,...,10), the energy of the system, $E(kT)$, can be expressed by means of the occupation probabilities

$$E(kT) = \sum_j \epsilon_j f_j(kT), \tag{9}$$

where

$$f_j(kT) = \sum_{n=1}^{N} P_{jn}, \quad P_{jn} = |\langle \phi_j \mid \psi_n(kT) \rangle|^2. \tag{10}$$

Transition probabilities $P_{jn}$ are just the squared moduli of the matrix elements of the unitary evolution operator $\hat{U}(kT,0)$.

The energy dispersion for the $N$-particle system, $\sigma_E^2 = \langle \hat{H}^2 \rangle - \langle \hat{H} \rangle^2$, calculated at the time $t = kT$, reads

$$\sigma_E^2 = \sum_j \epsilon_j^2 f_j - \sum_{n=1}^{N} (\sum_i \epsilon_i P_{in})^2, \tag{11}$$

where we have omitted the term due to correlations induced by antisymmetrization, i.e. the Fock term, $-\sum_{n\neq m} |\langle \psi_n \mid \hat{H}_0 \mid \psi_m \rangle|^2$. Note that with the specified initial conditions $\sigma_E^2(t=0) = 0$.

Similar calculations as for the cavities were performed also for the diffused Woods-Saxon potential [1]. The depth of the potential was taken to be $V_0 = -200$ MeV in order to prevent particles from escaping. Two diffuseness parameters were studied. The smaller one, equal to 0.2 fm, makes the Woods-Saxon well very much alike the finite-depth cavity. The larger value of 0.9 fm corresponds more closely to the diffuseness of the nuclear mean potential, but is still effectively smaller due to the unusual depth of the well. The radius parameter of the Woods-Saxon potential was fixed for each diffuseness value to keep the nominal Fermi energy of the $N$-particle gas constant [2].

## III. NUMERICAL CALCULATIONS

The solution of the time-dependent Schroedinger equation in a hard-walled cavity has been obtained by expanding the wave functions in a suitable basis [1].

One should remember that in spite of non-integrability of motion in cavities with $P_{n=3,4,5,6}$ deformations one still has one good quantum number $m$ associated with the projection of the angular momentum on the symmetry axis. Moreover, in the case of reflection symmetric deformations, parity is also conserved. Thus we have in general a number of non-communicating gases of fermions (see Ref. [1] for details). Since the $m=0$ levels are doubly degenerate and the $|m|>0$ levels are four-fold degenerate each gas is described by a small number of evolving wave-functions (e.g., for $P_3$ oscillation there are 12 levels in the subspace with $m=0$ but only one level with $|m|=5$ and 6).

Since we use a finite basis in our calculations there is a question of accuracy related to the truncation of the basis. Calculations were performed with the number of basis states corresponding to the cutoff for zeroes of spherical Bessel functions $j_l$, $a_n^l < 35$, which corresponds to 145 states with $m=0$, both parities. To study the effect of truncation we increased the cutoff to $a_n^l < 40$ which corresponds to 187 $m=0$ states with both parities. Due to the increase of the dimension of the basis, the excitation energy of the quantum gas after 10 periods for the largest studied $\eta=0.30$ changed by 2.6% for $P_3$ oscillations, and by 2.5% for $P_4$ oscillations. This, together with the smallness of the occupation probabilities of high-lying states, gives some confidence in our results.

The expansion of the wave-functions in the deformed harmonic oscillator basis was used when solving the time-dependent Schroedinger equation for particles in the diffused Woods-Saxon potential. The convergence of the method with the number of equivalent spherical oscillator shells, $N_0 = n_z + n_\perp$, taken into account was checked for the $P_4$ mode of oscillations and the numerically more demanding case of the small diffuseness of 0.2 fm. The relative excitation energy, $\Delta E/E_0$, after 10 cycles of oscillation with $\eta=0.3$ was equal to 1.5498, 1.7997 and 1.7394 for $N_0 =18$, 20 and 22, respectively. Therefore in all calculations presented below, the basis of $N_0 =20$ shells was adopted.

## IV. ENERGY DIFFUSION IN A CHAOTIC ADIABATIC BILLIARD GAS

In this section we review the classical description of a gas of particles in the container whose shape is slowly varying in time. We summarize some useful relations which will be compared with the quantal results. For derivations see Refs. [10,7].

Assuming the chaoticity and mixing property of the classical system evolving under slowly time-varying Hamiltonian one can arrive at the Fokker-Planck equation which governs the time evolution of the energy distribution of the ensemble of such systems. The usual assumptions made include that of the characteristic time scale of the collective motion much longer than the correlation time. The latter is the time over which a particle 'remembers' its velocity and is inversely proportional to the largest Lyapunov exponent.

This Fokker-Planck equation is equivalent to the following set of equations for moments of the velocity distribution [14]

$$\frac{d\langle v^n \rangle}{dt} = \frac{n(n+2)}{4} \langle v^{n-1} \rangle \frac{1}{V} \oint \dot{n}^2 d\sigma \ , \qquad (12)$$

where $V$ is the volume of the container, and $\dot{n}$ is the wall speed projected on the outward normal. The formula for $n = 2$ is just the wall formula of Ref. [6]. Integrating Eqs. (12) one obtains formulas for all moments in terms of the variable $\tau(t) = \frac{15}{8} \int_0^t \frac{1}{V\langle v \rangle_0} \oint \dot{n}^2 d\sigma$, where the numerical coefficient in front is kept to have notation consistent with Refs. [1,2]. In particular, we have

$$\begin{aligned}
\langle v^2 \rangle &= \langle v^2 \rangle_0 + \frac{16}{15}\langle v \rangle_0^2 (\tau + \frac{1}{5}\tau^2) \\
\langle v^4 \rangle &= \langle v^4 \rangle_0 + \frac{16}{5}\langle v \rangle_0 [\langle v^3 \rangle_0 \tau + \langle v \rangle_0 \langle v^2 \rangle_0 \tau^2 + \frac{16}{45}\langle v \rangle_0^3 (\tau^3 + \frac{1}{10}\tau^4)].
\end{aligned} \tag{13}$$

Equations (13) describe two effects. The first one is the increase of the mean energy of the particles. The second effect is the increase of the spread of the velocity distribution around the mean, which can be measured by the increase in the energy variance, $\sigma_E^2 - \sigma_{E0}^2 = \langle (E - \langle E \rangle)^2 \rangle - \langle (E - \langle E \rangle)^2 \rangle_0$. One can apply Eqs. (13) to an arbitrary initial velocity distribution function.

For one particle with sharply defined initial velocity, $\langle v^n \rangle_0 = v_0^n$, we obtain

$$\begin{aligned}
\frac{\Delta E}{E_0} = \frac{\langle E \rangle - \langle E \rangle_0}{\langle E \rangle_0} &= \frac{16}{15}[\tau + \frac{1}{5}\tau^2] \\
\frac{\sigma_E^2 - \sigma_{E0}^2}{\langle E \rangle_0^2} &= \frac{16}{15}\tau[1 + \frac{23}{15}\tau + \frac{16 \cdot 3}{15 \cdot 5}(\tau^2 + \frac{1}{10}\tau^3)].
\end{aligned} \tag{14}$$

Assuming the sharp Fermi distribution of initial velocities, $\langle v^n \rangle_0 = \frac{3}{n+3} v_F^n$, we obtain for the system containing one particle

$$\begin{aligned}
\frac{\Delta E}{E_0} = \frac{\langle E \rangle - \langle E \rangle_0}{\langle E \rangle_0} &= \tau + \frac{1}{5}\tau^2 \\
\frac{\sigma_E^2 - \sigma_{E0}^2}{\langle E \rangle_0^2} &= \tau[\frac{4}{3} + \frac{8}{5}\tau + \frac{3}{5}\tau^2 + \frac{3}{50}\tau^3].
\end{aligned} \tag{15}$$

The first identities in Eqs. (14, 15) remain true also for the system of $N$ particles. The right-hand side of each of the second identities expresses the quantity $N(\sigma_E^2 - \sigma_{E0}^2)/E_0^2$ for such system. The approximate expression for the variable $\tau$ in the case of vibrations around the sphere and the initial Fermi distribution of velocities reads

$$\tau = \frac{3}{4}c\eta[\omega t - \frac{1}{2}\sin 2\omega t], \tag{16}$$

with $\eta$ given by Eq. (6). Corrections to this expression, following from the exact treatment of the normal to the container surface, are small [2]. In the case of particles with all initial velocities equal to $v_0$, the appropriate expression for $\tau$ is multiplied by the additional factor $3/4$, with the adiabaticity parameter $\eta = c\omega R_0/v_0$. Since we are interested mainly in the average slope predicted by the classical expressions we will neglect the oscillating term in the $\tau$ function and calculate classical quantities replacing $\tau$ by $\tilde{\tau} = \frac{3}{4}c\eta\omega t$.

There have been efforts to improve the wall formula by allowing for special quantum corrections (see e.g. [2]). One improvement takes into account the suppression of the excitation due to the finite ratio of the particle wavelength to the wavelength of the multipole ripple of the oscillating cavity [15]. The other one corrects both the gas density and its average velocity for 'quantal squeezing' [2]. Since the main effects discussed in this paper turn out to be very pronounced while these two corrections act in opposite directions we did not use them in comparisons with the results of quantum calculations.

## V. RESULTS

The relative excitation energy of the gas in the oscillating cavity, $\Delta E/E_0$, as a function of time during 10 periods of oscillation is shown in Fig.1. The corresponding quantity for the gas in the Woods-Saxon potential with the diffuseness parameter of 0.9 fm is plotted in Fig.2. Different curves in each subfigure refer to different adiabaticity parameters $\eta$ (or oscillation frequencies $\hbar\omega$) and the order of curves, from the bottom to the top, corresponds to the increasing order of $\eta$: 0.06, 0.12, 0.18, 0.24, 0.30. In the most involved plot for the spheroidal oscillations of the Woods-Saxon potential, the three lowest curves may be identified by the correspondence of the points $\Delta E/E_0$ at $t/T = 1/2$, counted from the bottom up, to $\eta = 0.06, 0.12, 0.18$, respectively.

The results for the Woods-Saxon potential involve some non-trivial dependence on the diffuseness parameter, discussed in Ref. [2]. Hence the quantitative comparisons with the wall formula are more straightforward for a container. For this reason the rest of presented results concerns particles in the hard-walled cavity.

In Fig. 3, the comparison of the quantal results with the wall formula predictions has been shown for two values of the adiabaticity parameter, $\eta = 0.18$ and 0.30. In Figs.4 and 5, we have compared quantum results for the excitation energy, $\Delta E/E_0$, and the increase in energy variance, $N(\sigma_E^2 - \sigma_{E0}^2)/E_0^2$, after 10 periods of oscillation with classical predictions of Eq.(15), plotted as a function of the adiabaticity parameter. The latter quantity, Fig.5, measures the spread of the energy distribution in the quantum gas.

Another view on the excitation of the quantum gas is provided in Fig. 6, where we show the occupation probabilities $f_i$ of the $m = 0$ levels ($m = 0, \pi = +$, for the parity-preserving oscillations) after 10 periods for five types of cavities oscillating with $\eta = 0.3$. The transition probabilities $P_{in}$ from two selected states with $m = 0$ ($m = 0, \pi = +$), one close to the Fermi level and the second one with energy close to $3/5\epsilon_F$, are shown in Fig. 7. This latter figure provides some characteristics of the randomization property of quantum dynamics.

In Fig. 8, the actual increase in energy and in the energy dispersion for a single state in the cavity oscillating in the $P_6$ mode are compared to the classical estimates of Eq. (14). In Fig. 8a, the results for two different $\eta$ are shown for the particle occupying initially a level close to the Fermi surface. In Fig. 8b, the results for two orbits, one at the Fermi level and the other one at the bottom of the well, are shown. Here, one can appreciate different response of individual particles, sitting initially in different orbits, to the external perturbation.

## VI. DISCUSSION

The two most conspicuous features seen in Figures 1-6 are that 1) there is a qualitative difference in the energy evolution between the spheroidal vibrations and the rest, and 2) in all the cases studied, except for the $P_3$ oscillations of the Woods-Saxon potential with $\eta = 0.30$, the classical predictions largely overestimate the quantum results for times longer than approximately two periods.

The first effect is known from classical calculations [11] and was related to the integrability of motion in the spheroidal well. Since it was discussed in previous papers we make here only few comments.

The comparison of the excitation in spheroidal cavities with the wall formula after one period (see Refs. [1,2]) does not show entire qualitative difference between the two. This becomes evident only after longer times, as the initial increase in energy saturates for all $\eta$ and then (Fig.1) $\Delta E/E_0$ exhibits almost elastic (i.e. reversible) behaviour. Probably, at the beginning of the oscillation, particles inside a container do not 'realize' that the shape is integrable.

Additional confirmation of the speciality of the spheroidal oscillations comes from Figs.4 and 5. It is clear from the latter that not only the excitation energy but also the spread of energy distribution is nearly vanishing in the quantum case. Another representation of this fact is evident in Fig. 6, where occupation probabilities after 10 periods of spheroidal oscillations with $\eta = 0.3$ are seen to be confined to a small number of low-lying levels.

For other types of cavity oscillations the almost adiabatic motion takes place for $\eta = 0.06$ for reflection symmetric vibrations (see Fig.1). In contrast, for the $P_3$ oscillations, the steady increase of energy is seen even for the smallest value of $\eta$ studied. This effect can be understood as a consequence of the mixing of states of opposite parities ($\pi$), and hence the doubling of the effective level density, for the reflection asymmetric shapes as compared to the case of $P_4$ and $P_6$ deformations. However, this effect seems to be weaker for $P_5$ vibrations.

Considering much smaller excitation of the gas during $P_{3,4,5,6}$ oscillations than predicted by the wall formula it must be emphasized that this is a purely quantum effect. It is at variance with results of numerical simulations for the classical gas (see Fig.3 of Ref. [11]) that follow quite closely the wall formula predictions for $P_4$, $P_5$ and $P_6$ vibrations, and fall little below for the $P_3$ mode, however, still overshooting the linear part of Eq. (15). This discrepancy between classical and quantum results was not seen earlier [1], when only one period of oscillation has been studied. We suspect that it is related to the periodicity of shape changes which imposes additional correlations on the quantum motion of particles. Note that the qualitative change of the excitation energy curve is seen around the second period of oscillations when the particle can 'realize' that the vibrations are periodic.

The excitation energies of the gas in the Woods-Saxon potential with the diffuseness 0.2 fm (not shown) are qualitatively similar to those for the cavity with some quantitative differences. The excitation energies are smaller for the $P_3$ mode and larger for the rest, especially for the $P_5$ vibration with $\eta = 0.24, 0.30$, for which they increase roughly by 50%. The results for the Woods-Saxon well with the diffuseness 0.9 fm, Fig.2, also show, except for $P_4$, some increase in excitation as compared to Fig.1. The change in $\Delta E/E_0$ is quite substantial for $P_6$ oscillations but the most dramatic increase, roughly by a factor of 2.5, occurs for the $P_3$ mode with $\eta =0.24$ and 0.30. The $P_3$ vibration with $\eta = 0.30$ is the only case in which $\Delta E/E_0$ does not fall below the wall formula value up to $t/T=5$. Another difference between the results for the cavity and the Woods-Saxon potential concerns the wiggles in the excitation energy plot, which are somewhat smoothed out in Fig.2 as compared to Fig.1. All these changes of $\Delta E/E_0$ with the diffuseness of the well, which follow the trends shown in Fig.3 of Ref. [2], do not affect the main conclusion following from Figs.1 and 2: The excitation rate of the quantum gas becomes suppressed in many-cycle oscillations. Initial slope of $\Delta E/E_0$ as a function of time is clearly reduced after few (2-5) oscillation cycles, independent of the special form of the potential.

One can observe in Figs. 4, 5 that all quantum results, considered as functions of $\eta$, lie systematically below the classical ones. They also show some irregularities that can be attributed to the interference between subsequent Landau-Zener transitions. It should be noted that during one period of oscillation every avoided crossing is traversed twice. Hence the resulting transition probability after one period is not merely the product of asymptotic Landau-Zener probabilities but is strongly affected by the phase relations which can enhance or hinder the resulting transition probability [16]. The effect will vanish if many transitions are involved with random phase relations [17], but this is not the case in our calculations, where usually only few avoided crossings near the Fermi level play a role.

The prediction for the excitation energy of the externally driven quantum system governed by the Hamiltonian being a random matrix has been given in Ref. [12]. It was presented as the second moment of some probability function $P(n)$, $\Delta E = \frac{1}{2\rho_F} \int (n - N_F)^2 P(n - N_F) dn$. In this formula, $\rho_F$ is the level density at the Fermi surface, and $P(n - N_F)$ is the probability of the transition from the state at the Fermi level $N_F$ to any other state $n$ during the oscillation time (averaged over some energy interval around the state $N_F$). The probability $P(n - N_F)$, assumed in [12] to be a function of the difference of level numbers only, corresponds to the quantities $P_{nN_F}$, introduced in section II, which, however, depend in fact on two levels. The important point in derivations presented in [12] was that the probability $P(n)$ fulfiled the diffusion equation, i.e. it took a gaussian shape with the width increasing linearly with time. In Fig.7, we show $P_{in}$ following from our quantum calculations for two states $n$ from the group of levels with $m = 0$ or $m = 0, \pi = +$: one at the Fermi level and the other one at the 'average' energy $\epsilon \approx 3/5\epsilon_F$. It is clearly seen in Fig.7, that neither of them is a gaussian. In fact, it is also seen that our systems, although being classically chaotic (but not completely chaotic) preserve some amount of selection rules for transitions. Indeed, transitions to some states are much more probable than to the neighbouring ones (Fig. 7). The effect can also be seen in Fig.6 where the occupation probabilities $f_i$, plotted after ten periods, exhibit the irregular structure. This is another illustration of the fact that one cannot treat the gases considered in this paper by means of a theory assuming one of the universality classes of random matrix ensembles. It turns out that the excitation energy of particles with $m = 0$, calculated according to the prescription given in [12], overshoots the quantum results in a similar manner as the wall formula does.

The effect of saturation of the excitation in the fully chaotic system being subject to the periodic perturbation has been reported in Ref. [12], where it was related to the localization of the Floquet operator eigenstates. Although the effect of saturation can be visible after the number of periods of the order of the number of Floquet states that form the initial state, which could be a relatively large number, the divergence from the expected linear increase of excitation can be visible even after much smaller number of periods (see [12]). On the basis of results shown in Figs. 1-6 we cannot claim that we see the effect of saturation of excitation in our case (except for spheroidal vibration, where it has classical origin). What we observe may be the early signature of it.

At the end of this section we address the question of averaging effects in quantum calculations. The excitation energy is the sum of single-particle contributions in the classical as well as in the quantum case. The actual adiabaticity parameter is very different for, say, the lowest state and for the state at the Fermi surface. This implies a different behaviour of particles occupying states of very different energies. Are these differences in the quantum case similar to those in the classical case ? We have chosen the $P_6$ vibrations to illustrate this problem.

In Fig. 8a, we have shown the comparison of the calculated $\Delta E/E_0$ with the wall formula (upper panel) for one state with angular momentum projection $m = 0$ and parity $\pi = +$, located close to the Fermi level. In order to eliminate effects associated with changes of the deformation speed and the adiabatic energy during one cycle, only the points corresponding to the integer number of cycles have been plotted. The adiabaticity parameters $\eta = 0.18$ and 0.3 are included. A relatively good agreement with the wall formula can be claimed only for the first period of oscillation for both values of $\eta$.

In the lower panel, we show the behaviour of the second cumulant of energy for the same state. One can notice suprisingly good agreement with the classical prediction in both cases. Note that even for times, when the wall formula predictions fail, the diffusion process is still well described by the classical expression. This picture, however, changes if we consider the lower energy states.

In Fig. 8b (upper panel), the quantum excitation energy and its classical estimate for $\eta = 0.30$ have been plotted for the same state at the Fermi level as in Fig.8a and the other one, with $m = 0$ and $\pi = +$, at the bottom of the potential. Note that the lowest state does not show any excitation, in contrast to classical predictions (dashed-dotted line with full circles). This happens in spite of the fact that its actual adiabaticity parameter, $\eta_{ac} = c\omega R_0/v_0 \approx 3\eta$, is close to 1, i.e. the surface vibration is fast for this particle. The lower panel of Fig.8b shows second cumulants of energy. Again, one can see the striking disagreement with classical predictions for the lowest state in the well.

One can venture to identify the reason of this disagreement between quantum and classical results for low-lying states with the suppression of excitation for particle wavelengths larger than that of the ripple of the cavity wall [15]. Indeed, for oscillations of multipolarity $n$ the ratio of the Fermi wavelength to the wavelength of the wall oscillations equals $0.1081n$ [1]. For a particle in the lowest level this ratio is larger by the factor $\eta_{ac}/\eta = 1/0.3$, i.e. it equals 2.162 for $n = 6$. For the wavelengths ratio larger than 2 the theory presented in [15] predicts the complete suppression of excitation.

## VII. CONCLUSIONS

In the present paper we have reported on the divergence of the excitation energy of a quantum gas, driven by periodic oscillations of the potential well, from the classical predictions. The five types of deformations were considered: spheroidal and Legendre polynomial ripples: $P_3$, $P_4$, $P_5$ and $P_6$. The excitation energy was followed in time for ten oscillations for five different values of the adiabaticity parameter. We have found qualitative differences between the wall formula predictions and quantum results for nonintegrable hard-walled cavities: the suppresion of the excitation and of the spread of energy distribution, and the remnant selection rules seen in transition probabilities from particular states. The suppression of excitation seems to be quite universal phenomenon as we obtained it also for the gas in the diffused Woods-Saxon well. Most likely, one can attribute these effects to the periodicity of driving.

In view of these results it seems that the use of the wall formula for a description of nuclear dissipation in cyclic dynamical processes (like shape oscillations) may be questionable. This would be an another obstacle in using classically motivated formulas, which before were found inappropriate for the class of integrable shapes of the cavity.

Moreover, we find that while the wall formula predictions are in tolerable agreement with the time evolution of particles near the Fermi surface during first few oscillation periods, the particles placed initially in the lowest states show small excitation or do not excite at all. We connect this fact with the quantum suppression of excitation for particles with the wavelength larger than the wavelength of the cavity ripples [15].

## ACKNOWLEDGMENTS

The authors thank W.J. Swiatecki for numerous enlightening discussions. This research was supported in part by the Polish Committee for Scientific Research under Contract No. 2 P03B 034 08, and by a computational grant from the Interdisciplinary Centre for Mathematical and Computational Modeling (ICM) of Warsaw University. Two of the authors (J.S and J.B.) benefitted from the hospitality of the Theory Group in the Nuclear Science Division at the Lawrence Berkeley Laboratory on the occasion of several visits. The access to computing facilities at the LBL is very much appreciated. The financial support by the Polish-American Maria Sklodowska-Curie Joint Fund II, under project PAA/NSF-96-253, is gratefully acknowledged.

FIGURES

FIG. 1. The relative excitation energy, $\Delta E/E_0$, of 112 fermions in five types of cavities oscillating around the spherical shape, plotted as a function of oscillation cycles. Different curves in each subfigure, from the bottom to the top, correspond to the adiabaticity parameters $\eta = 0.06, 0.12, 0.18, 0.24, 0.30$, respectively.

FIG. 2. As Fig.1 but for the Woods-Saxon potential with the diffuseness 0.9 fm. For details see text.

FIG. 3. The relative excitation energy, $\Delta E/E_0$, of 112 fermions in four nonintegrable cavities oscillating around the spherical shape, plotted as a function of oscillation cycles for two values of $\eta$ (0.30 – dot-dashed line and 0.18 – dashed line). The wall formula predictions are denoted by the respective curves with dots.

FIG. 4. The relative excitation energy, $\Delta E/E_0$, of 112 fermions after 10 periods of oscillation in five types of cavities, plotted as a function of the adiabaticity parameter (full circles). The wall formula predictions are also shown (solid line).

FIG. 5. The energy variance of the gas in five types of cavities after 10 oscillation periods, plotted as a function of the adiabaticity parameter (full circles). The predictions according to Eq.(15) are also shown (solid line).

FIG. 6. The occupation probabilities $f_i$ for $m = 0$ levels ($m = 0$, $\pi = +$, for parity-preserving vibrations) after ten periods of oscillation for $\eta = 0.30$.

FIG. 7. The transition probabilities $P_{in}$ (see text for definition) from two states placed at $\epsilon \approx \epsilon_F$ and $\epsilon \approx 3/5\epsilon_F$ for four types of cavities and $\eta = 0.3$ ($\sum_i P_{in} = 1$).

FIG. 8. The excitation energy (upper panels) and energy variance (lower panels) for one particle with $m = 0$ and $\pi = +$ in the cavity oscillating in the $P_6$ mode (a) - $\eta = 0.18$ and 0.30, the particle occupies initially the state close to the Fermi level; (b) - $\eta = 0.3$, the particle occupies initially the state at the Fermi level (solid line) or at the bottom of the well (dot-dashed line). The classical predictions are denoted by respective lines with full circles.

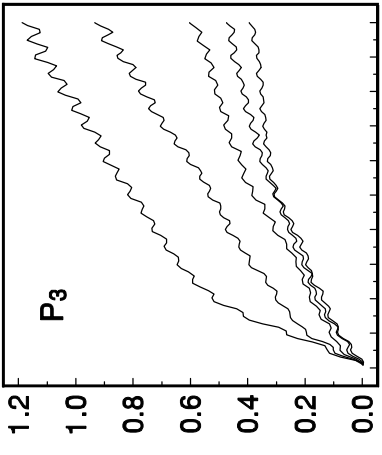
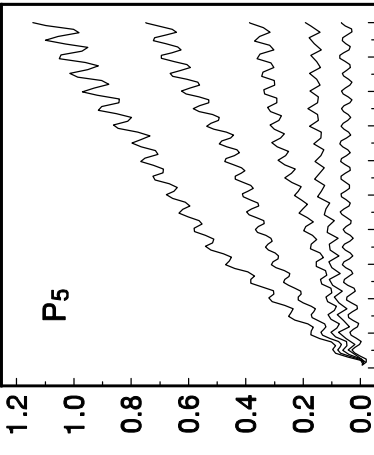
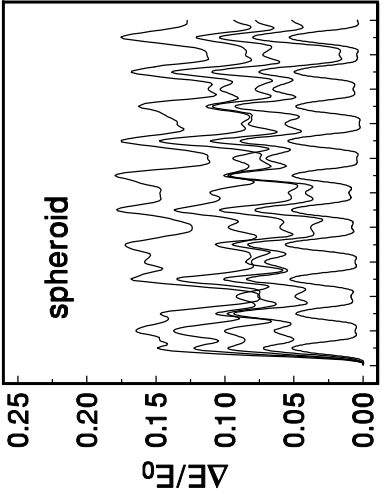
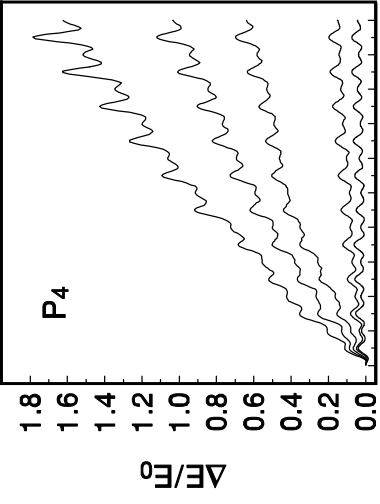
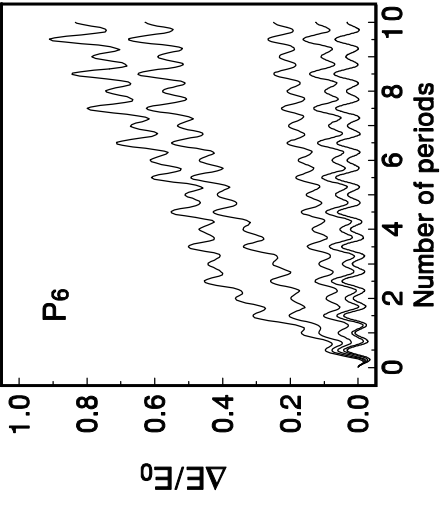

$V_0 = -200$ MeV dif=0.9fm $c_0 = 0$ $c_1 = 0.2$

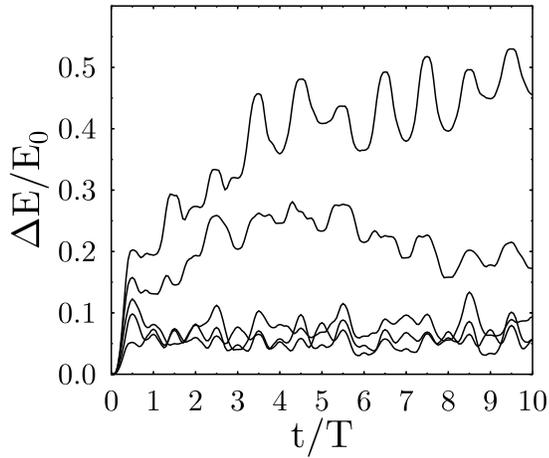
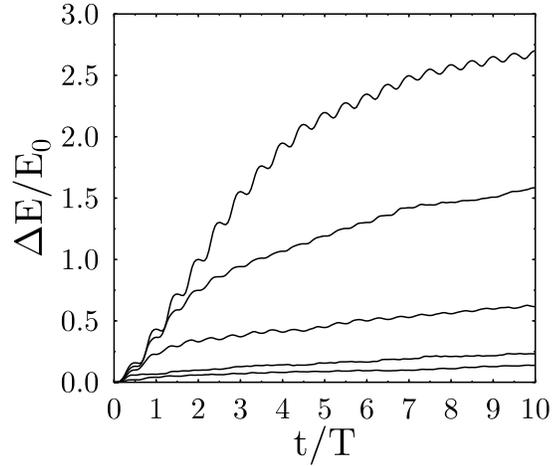
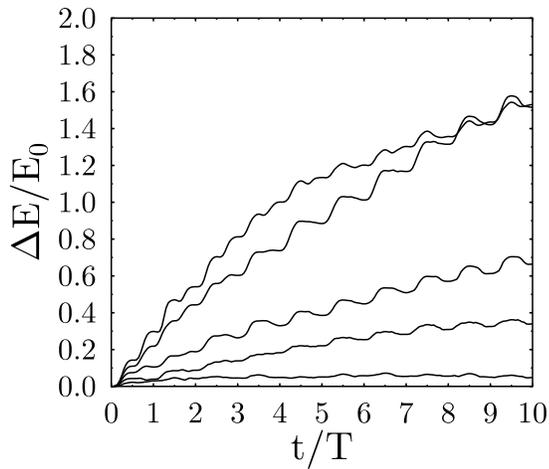
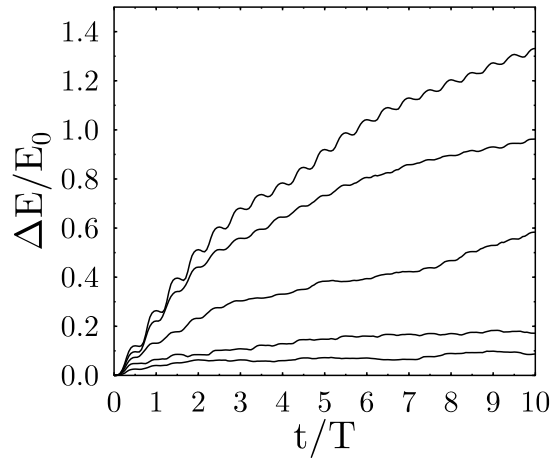
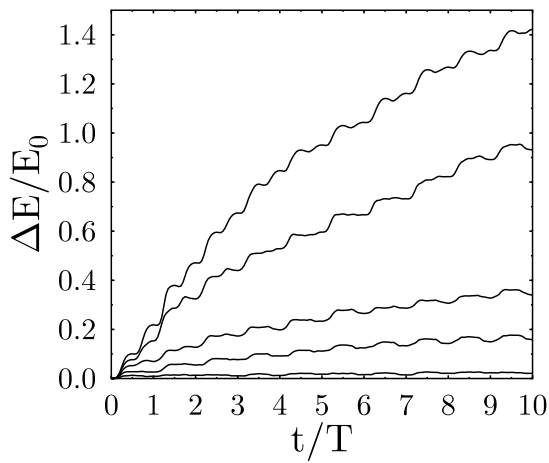

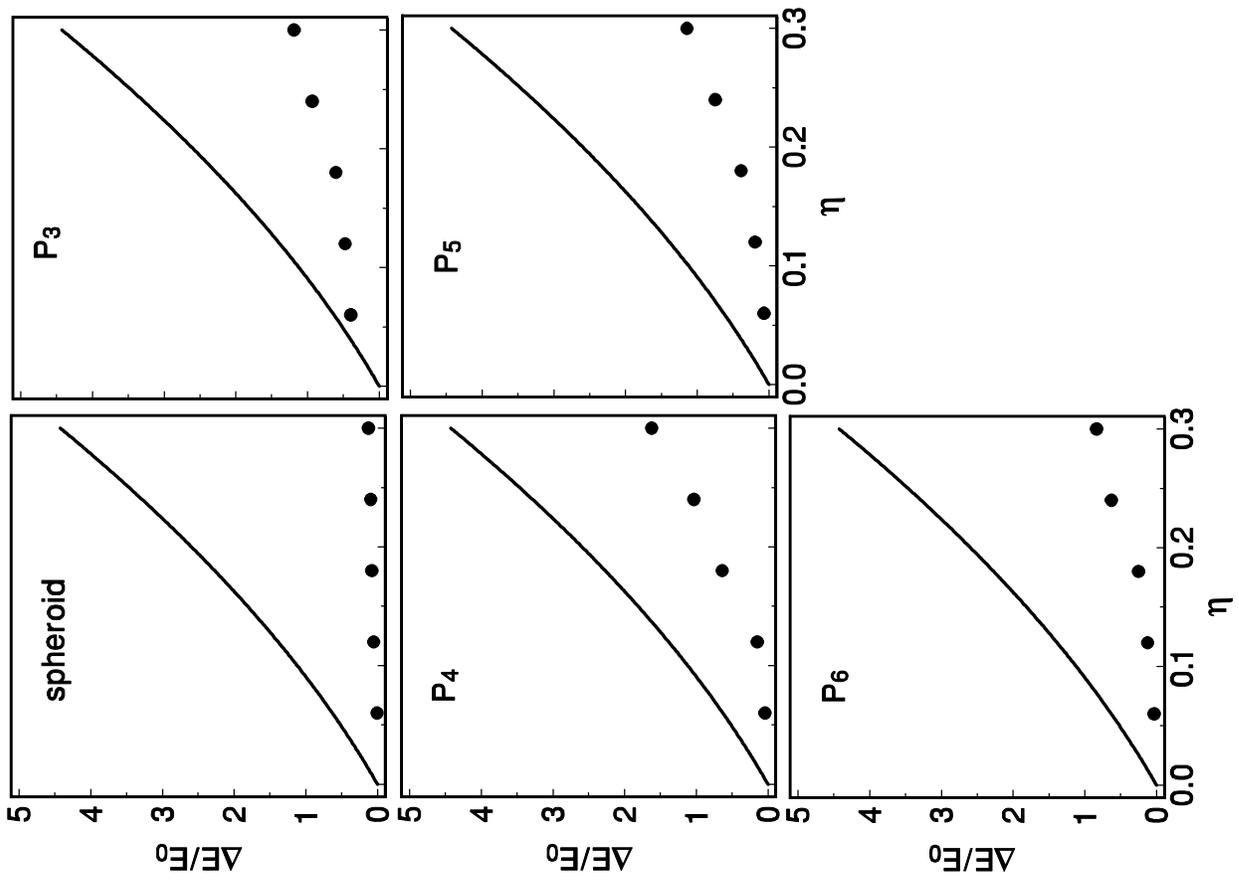

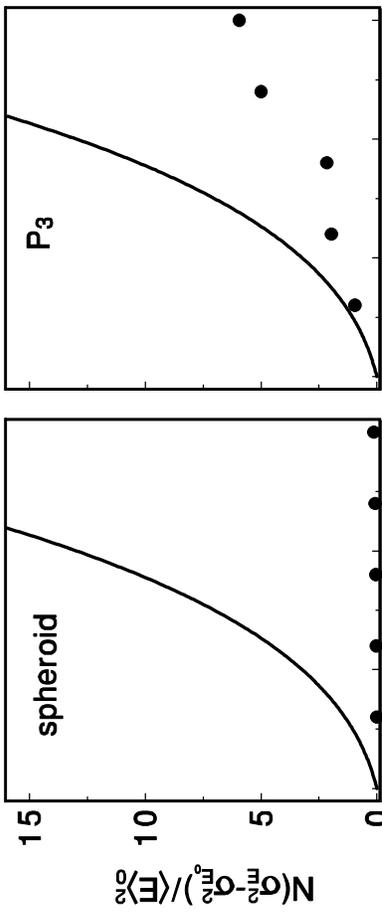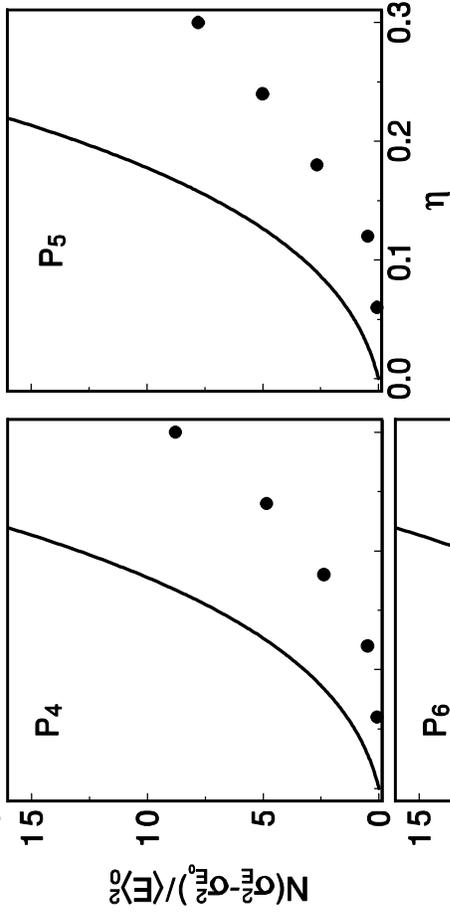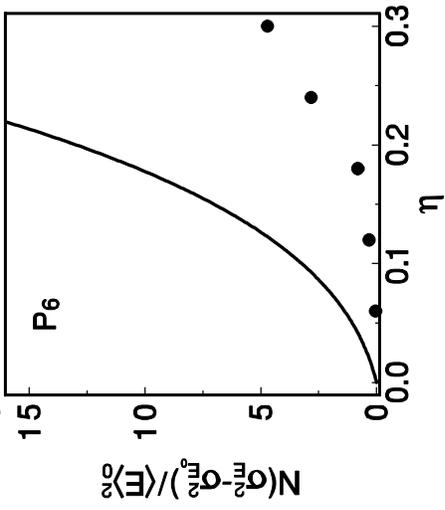

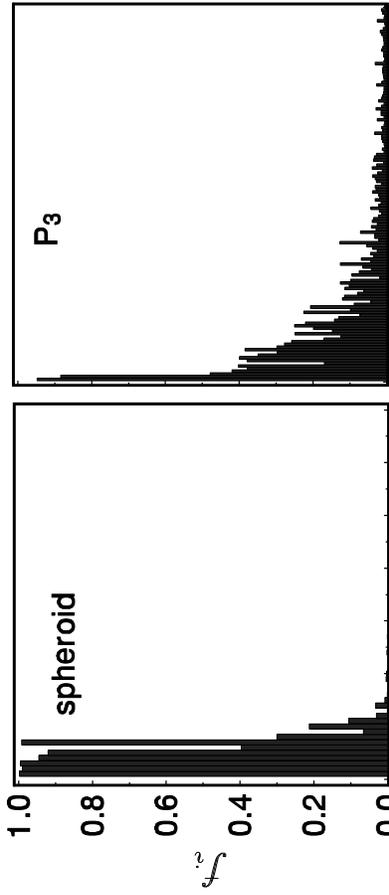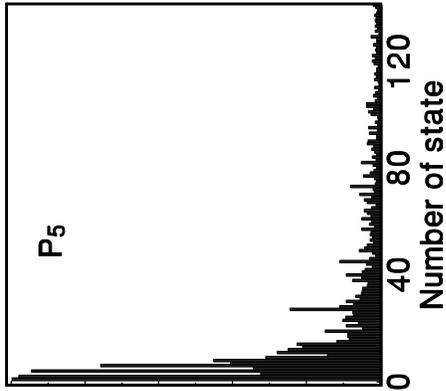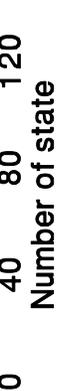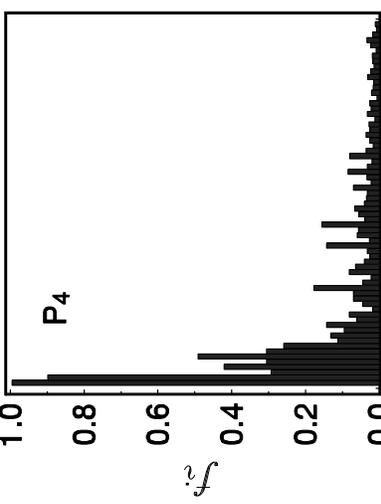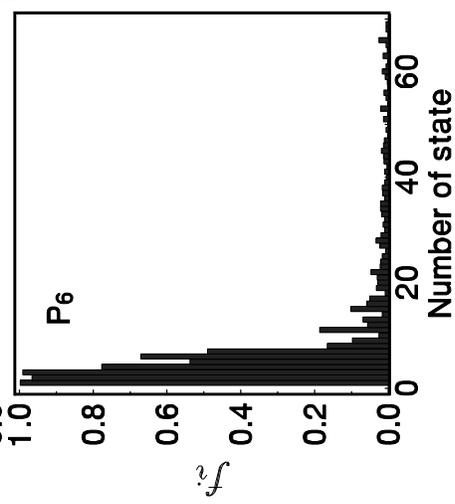

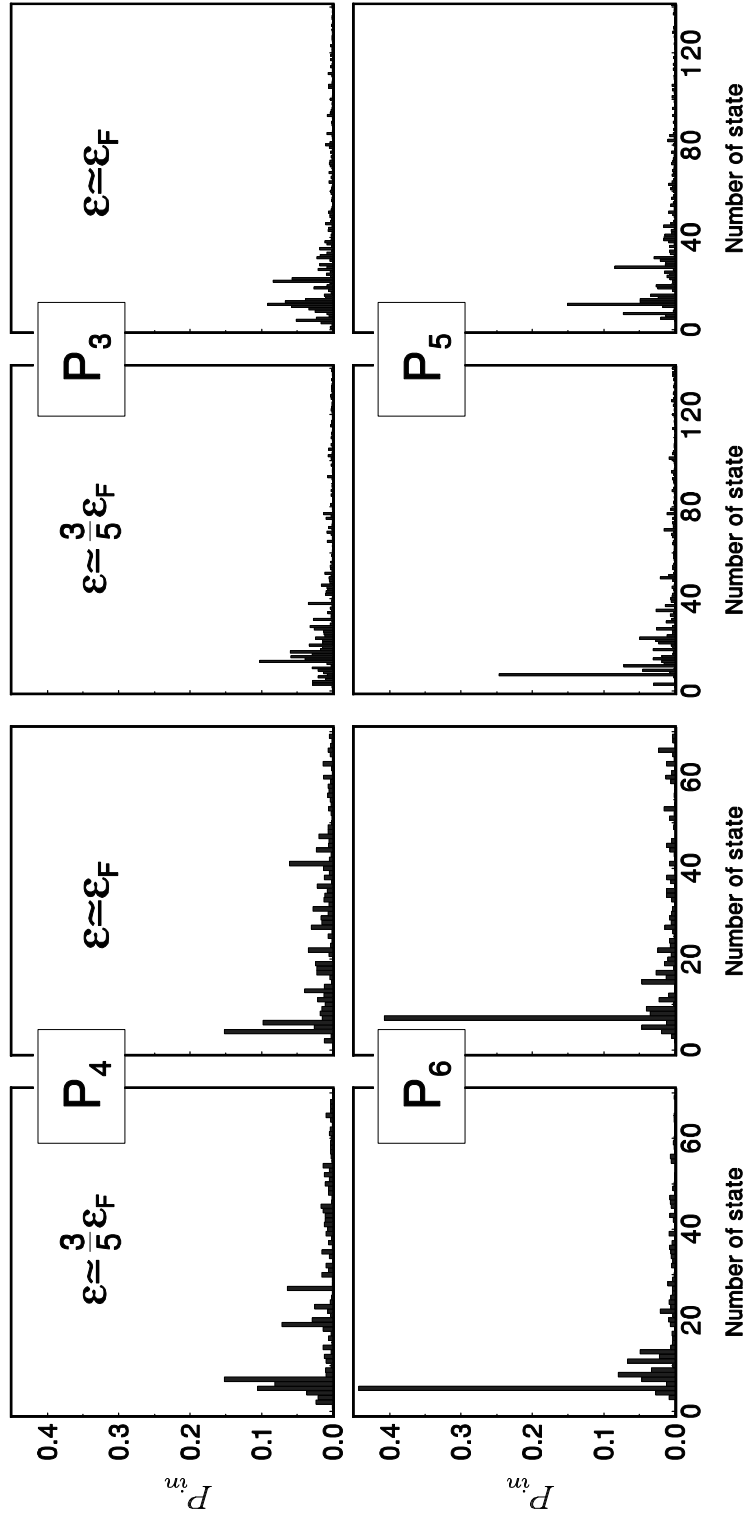

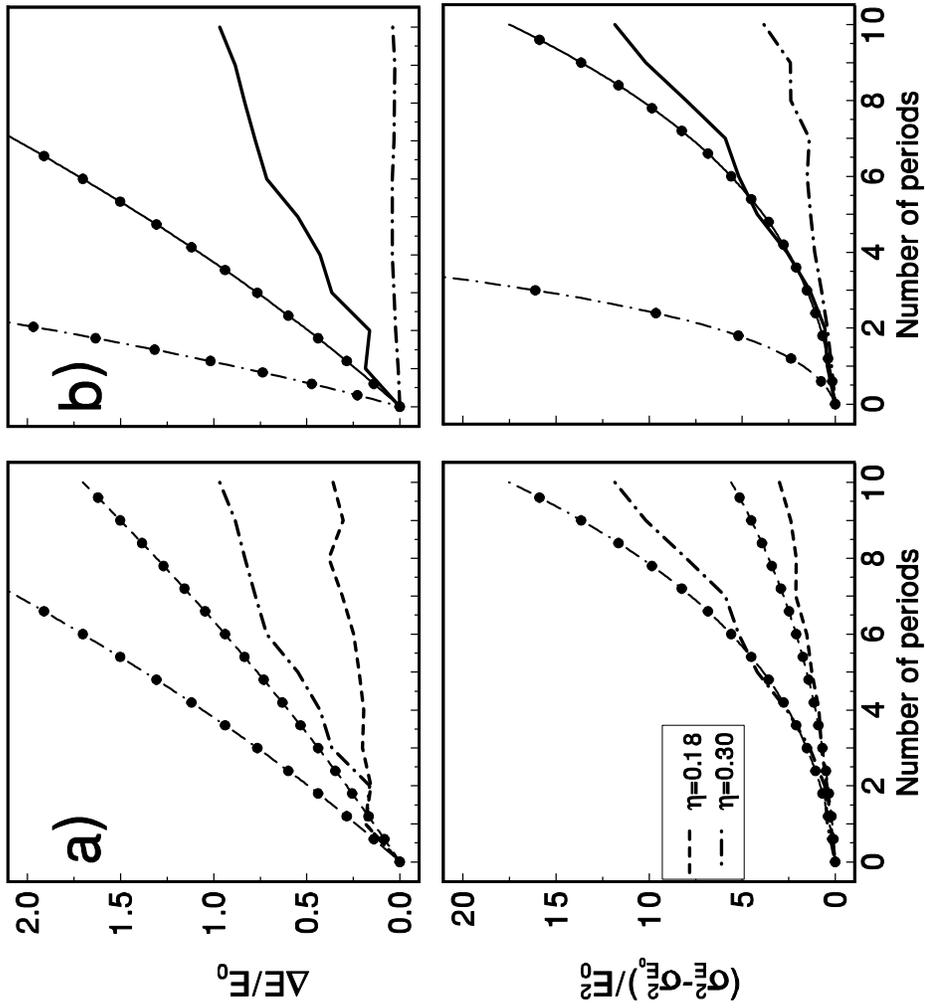